\documentclass[twocolumn]{aa}
\usepackage{graphicx,subfigure}
\usepackage{txfonts}
\begin{document}
   \title{Combining visibilities from the Giant Meterwave Radio Telescope
and the Nancay Radio Heliograph}

   \subtitle{High dynamic range snapshot 
images of the solar corona at 327 MHz}

   \author{C. Mercier\thanks{claude.mercier@obspm.fr},
          \inst{1}
    Prasad Subramanian,\inst{2}Alain Kerdraon,\inst{1} Monique Pick,\inst{1} 
          S. Ananthakrishnan,\inst{3} P. Janardhan\inst{4}
          }

   \offprints{C. Mercier}

\institute{Observatoire de Paris, LESIA, Groupe RH, 5 place Janssen 92190 
	Meudon, 
   France\\
\and
   Inter-University Centre for Astronomy and Astrophysics, P. O. Bag 4, 
	Ganeshkhind, Pune-411007, India\\
\and
    NCRA-TIFR, P. O. Bag 3, Ganeshkhind, Pune University Campus, Pune-411007, 
	India\\
\and
    Physical Research Laboratory, Astronomy \& Astrophysics Division, 
	Navrangpura, Ahmedabad-380009, India
             }
             
   \date{}

   \abstract {
We report first results from an ongoing program of combining visibilities 
from the Giant Meterwave Radio Telescope ($GMRT$) and the Nancay Radio 
Heliograph ($NRH$) to produce composite snapshot images of the sun at meter 
wavelengths. We describe the data processing, including a specific multi-scale
$CLEAN$ algorithm. We present results of a) simulations for two models of the
sun at 327 MHz, with differing complexity b) observations of a complex noise 
storm on the sun at 327 MHz on Aug 27 2002. Our results illustrate the 
capacity of this method to produce high dynamic range snapshot images when 
the solar corona has structures with scales ranging from the image resolution 
of 49$^{''}$ to the size of the whole sun. 

We find that we cannot obtain reliable snapshot images for complex objects 
when the visibilities are sparsely sampled. 

\keywords{radioastronomy, high dynamic imaging, sun}
   }
\authorrunning{Mercier et al.}
\titlerunning{Combining visibilities from GMRT and NRH: High dynamic range 
   327 MHz snapshots}
\maketitle
%

\section{Introduction}

We have an ongoing program of combining data from the $GMRT$ and $NRH$ in order
to make composite meter wavelength images of the sun. Our aim is to combine
the short baseline data from the $NRH$ with long baseline data from the $GMRT$
in order to obtain meter wavelength images of unprecedented resolution and 
fidelity. We present here the first results from our campaign.

Imaging complex, rapidly varying radio sources in the solar corona is 
problematic because one needs to obtain images spanning durations as short as 
a few seconds. The uv-coverage for short duration exposures is usually quite 
poor, and this affects the ability to image complex sources reliably. 
Furthermore, deconvolution procedures like the commonly used $CLEAN$ often 
perform poorly on sources that exhibit structures on a variety of scales 
(Wakker \& Schwarz 1991). The problem of poor $uv$-coverage can be alleviated
by synthesis imaging over time durations of a few hours, but this technique 
cannot be used for studying phenomena that vary rapidly in time. There are 
several instances in solar physics where high dynamic range, high resolution 
snapshot images are essential. For instance, studies of extended thermal 
emission in the corona, coronal mass ejections (CMEs) that frequently occur 
together with compact, bright bursts and flare continua require imaging with a 
large field of view, high spatial resolution and dynamic range. High 
resolution observations can establish a firm limit on the angular broadening 
of sources in the solar corona due to turbulence, which determines the 
apparent size of radio bursts (e.g. Bastian 1994).
   
\subsection {Dynamic Range}   
Before proceeding further, we pause to define two kinds of dynamic ranges we 
will refer to below. The $rms$ dynamic range is the definition usually 
employed by the community. It is the ratio of the peak brightness of the image
to the $rms$ brightness of a representative empty region near the area where 
the peak brightness occurs. This definition is relevant when artefacts on the
final images look like random noise, which is often the case for images
produced via aperture synthesis. However, we deal here with snapshot images, 
for which the visibilities are rather sparsely sampled, especially for large
baselines. The artefacts on such snapshot images are localized and structured,
and often look like spurious compact sources. We will use another definition 
of dynamic range, which we call the ``$max$'' dynamic range, which we think is
more relevant to the snapshot images we present. The $max$ dynamic range is 
defined as $num/denom$, where $num$ is the peak brightness of the image. The 
quantity $denom$ is the maximum deviation from the mean of an empty region 
around the area where the peak brightness occurs. We will quote the $rms$ 
dynamic range for the sake of comparison with published results. The $max$ 
dynamic range is typically an order of magnitude lower than the $rms$ dynamic 
range for our results.

\subsection {Meter wavelength observations of the solar corona: a brief 
overview} 
Meter wavelength observations of the solar corona to date have concentrated 
either on having a large field of view with dense short baseline coverage 
(e.g., observations using the $NRH$) or on achieving high spatial resolution 
at the expense of a sparse short baseline coverage (e.g., extended array 
observations with the Very Large Array ($VLA$)). In what follows, we attempt 
to give a short overview of meter wavelength imaging observations of the solar 
corona. This will help to place the observations reported here in context.
    
There have been several attempts at detecting small scale structures at meter 
wavelengths in the solar corona. Kerdraon (1979) used one-dimensional ($1D$) 
scans at a time cadence of 0.02s and a spatial resolution of  $1.2^{'}$  with 
the $NRH$ at 169 MHz. The $1D$ scans were obtained by a simple Fourier 
transform ($FT$) of the observed complex visibilities ($CVs$). The field of 
view was around 1 degree, and the artefacts on the $1D$ scans were due to 
sidelobes of the response function. The ``$max$'' dynamic ranges of these 
$1D$ scans were 5--10. Kerdraon concluded that type 1 bursts at 169 MHz 
range in size from $0.7^{'}$ to $3^{'}$.  

Zlobec et al (1992) have used the VLA in its A configuration with a resolution
of 4$^{''}$ at 327 MHz to search for small scale structures in snapshot maps 
with a time cadence of 1.6 seconds. They find that the smallest type I source 
sizes they can observe are around 30$^{''}$. The largest structure that can 
possibly be imaged with the VLA A-array is around 110$^{''}$. However, owing 
to the shortage of short baseline spacings, Zlobec et al. (1992) mention that 
they cannot be confident about large scale structures. In some cases, they 
cannot be confident about the positions of the small structures. Furthermore, 
the $rms$ dynamic range of the images they obtain is around 10, at best.

Some other instances of snapshot imaging of noise storm sources in the solar 
corona at meter wavelengths are Willson et al. (1997); (VLA C configuration, 
resolution of 40$^{''}$ at 327 MHz, time cadence 3.3 seconds), Willson et al. 
(1998); (VLA A configuration, resolution 30$^{''}$ at 75 MHz, 5$^{''}$ at 327 
MHz, time cadence 10 seconds, source size 1.5$^{'}$ at 327 MHz, 2.5 $\times$ 
5$^{'}$ at 75 MHz), Willson (2000) (VLA D configuration, time cadence 3.3 
seconds, resolution 180$^{''}$ at 327 MHz).

There have also been several instances of synthesis imaging of noise storm 
emitting regions at meter wavelengths. Habbal et al. (1996) have used the VLA 
in its B configuration to observe noise storm sources with synthesis durations
ranging from around 100 mins to 3.5 hours. Their resolution was around 
94$^{''}$ at 327 MHz. Earlier, Habbal et al. (1989) had used the VLA in its 
C configuration to image a noise storm source at 327 MHz with a synthesis 
duration of around 4 hours. They were able to image structures ranging from 
around 80$^{''}$ to 240$^{''}$ with a resolution of 57$^{''} \times$ 47$^{''}$
at 327 MHz. Willson et al. (1997) have used data from the VLA C array to 
obtain 4 hour synthesis images of a faint transequatorial arch measuring 
600$^{''}$ at 327 MHz. In a multiwavelength study aimed at high dynamic range 
imaging, White et al. (1992) have used the VLA D configuration to obtain 4 
hour synthesis images of active region plages measuring around 600$^{''}$ 
with a resolution of around 200$^{''}$ at 327 MHz. The $rms$ dynamic range of 
their 327 MHz images was between 60 and 90. In a similar study at 1420 MHz, 
Gopalswamy et al. (1991) used the VLA D configuration to obtain 6-7 hour 
synthesis images of active region plages ranging from 50$^{''}$ to 200$^{''}$ 
with a resolution of 50$^{''}$. The $rms$ dynamic range of their images varies 
from 170 to 346.      

\subsection {Present study}
In this paper, we present composite snapshot images by combining visibilities 
from the $GMRT$ and the $NRH$ for :
\begin {itemize}
\item
simulations with two models of the radio sun at 327 MHz. The first model is
relatively simple and similar in many aspects to the real case reported in 
this paper. The second model is more complex, and we use this in order to 
investigate the limits of what we can hope for with snapshot imaging.
\item
observations of complex and evolving noise storm emitting regions in the solar
corona at 327 MHz on Aug 27 2002. These observations use an integration time
of 17 seconds. They illustrate the capabilities of the composite instrument
with real data, but since solar bursts vary over time scales that are much 
shorter than 17 sec., they are of limited use in arriving at astrophysical 
results.
\end {itemize}
This procedure complements the dense short baseline coverage of the $NRH$ with 
the long baseline coverage of the $GMRT$. We obtain snapshot maps of structures
with sizes ranging between the resolution (49$^{''}$) and the size of the
whole sun with unprecedented ($rms$) dynamic ranges of 250--420. 


\section {The Instruments}
The Nancay RadioHeliograph ($NRH$) and the Giant Meterwave Radio Telescope 
($GMRT$) both operate at meter wavelengths, and we currently have common 
observations of the solar corona at 236 and 327 MHz. The two instruments have 
up to 4 hours of common observing time each day. The following subsections 
give brief overviews of the two instruments.

\subsection {The Nancay Radio Heliograph ($NRH$)}
The $NRH$ (Kerdraon \& Delouis 1997) is located about 200 $km$ south of Paris 
(latitude $47^{o}\,23^{'}N$, longitude $2^{o}\,12^{'}E$, altitude 150 $m$). It 
consists of 44 antennas of size ranging from 2--10 $m$, spread over 
two arms ($EW$ and $NS$) with respective lengths of 3200 $m$ and 2440 $m$. The
EW and NS baselines $d_{EW}$ and $d_{NS}$ are from 0 to 3200 $m$ (resp 
2440 $m$) by steps 100 $m$ (resp 54 $m$). After January 2003 the number of 
antennas were raised to 48 and the EW minimum baseline is reduced to 50 $m$. 
The $NRH$ observes the sun 7 $h$ per day at up to 10 frequencies between 150 
and 450 MHz. The standard observing mode consists of Stokes I and V 
images at 164, 236, 327, 410 and 432 MHz each 150 msec. The resolution 
of two dimensional ($2D$) images and the field of view depend on the 
frequency and the season. For the observation reported here they are 
$\sim 4.8^{'}\times 3.0^{'}$ and $\sim 40^{'} \times 70^{'}$ respectively.

\subsection {The Giant Metrewave Radio Telescope ($GMRT$)}
The $GMRT$ is located about 80 km north of the city of Pune in Maharashtra, 
western India (latitude 19$^{o}\,06^{'}$N, longitude 74$^{o}\,03^{'}$E, 
altitude 650 m). It consists of thirty 45 $m$ diameter antennas spread over 
25 $km$. Half of these are in a compact, randomly distributed array of about 
1 square $km$ and the rest are spread out in an approximate `Y' configuration.
The $GMRT$ currently operates in the frequency bands 
around 235, 325, 610 and 1000--1450 MHz, with bandwidths ranging
from 16 to 32 MHz. Of these, the 235 and 327 MHz frequencies are most 
commonly used for solar observations. The shortest baseline is 100 m and 
the longest one is 26 km. Data is usually acquired at a time cadence of 16 
seconds, but it is presently possible to acquire data at intervals as short as
2 seconds. The automatic level control circuitry in the $GMRT$ data processing 
electronics has a response time of around 1 second. This limits the shortest 
integration time to be $\approx 2 sec$, which is still substantially longer 
than what can be achieved with the $NRH$. Solar observations are typically 
carried out from around 03:30 $UT$ to 12:30 $UT$. During such observations, 
the sun is typically observed for around 15--20 minutes, after which a nearby 
phase calibrator is observed for around 10 minutes. This cycle is then 
repeated. During a particular cycle, the antennas track the sun at a linear 
rate in $RA$ and $DEC$. At the beginning of the next cycle, the antennas are 
repositioned to the $RA$ and $DEC$ of the sun corresponding to the time when 
the cycle begins, and they then track the sun as before.
Further details about the $GMRT$ can be found at the URL
http://www.gmrt.ncra.tifr.res.in and in Ananthakrishnan \& Rao (2002), 
Swarup et al. (1991).

\subsection {The composite instrument}
It is useful to discuss at this point some properties of the composite 
instrument, for a better understanding of the subsequent data processing.
Fig. 1 shows the uv coverage of the joint instrument for 2002 aug 27 at 
09:30 $UT$, with baselines up to 12 000 $\lambda$ :
\begin{figure}
\centering
\subfigure[]{
\includegraphics[width=3.85in]{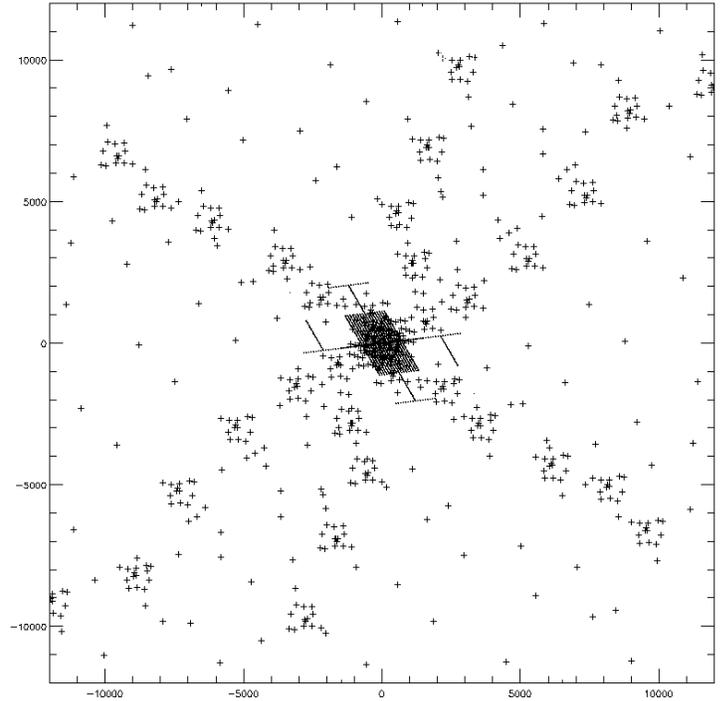}}
\hspace{1in}
\subfigure[]{
\includegraphics[width=3.85in]{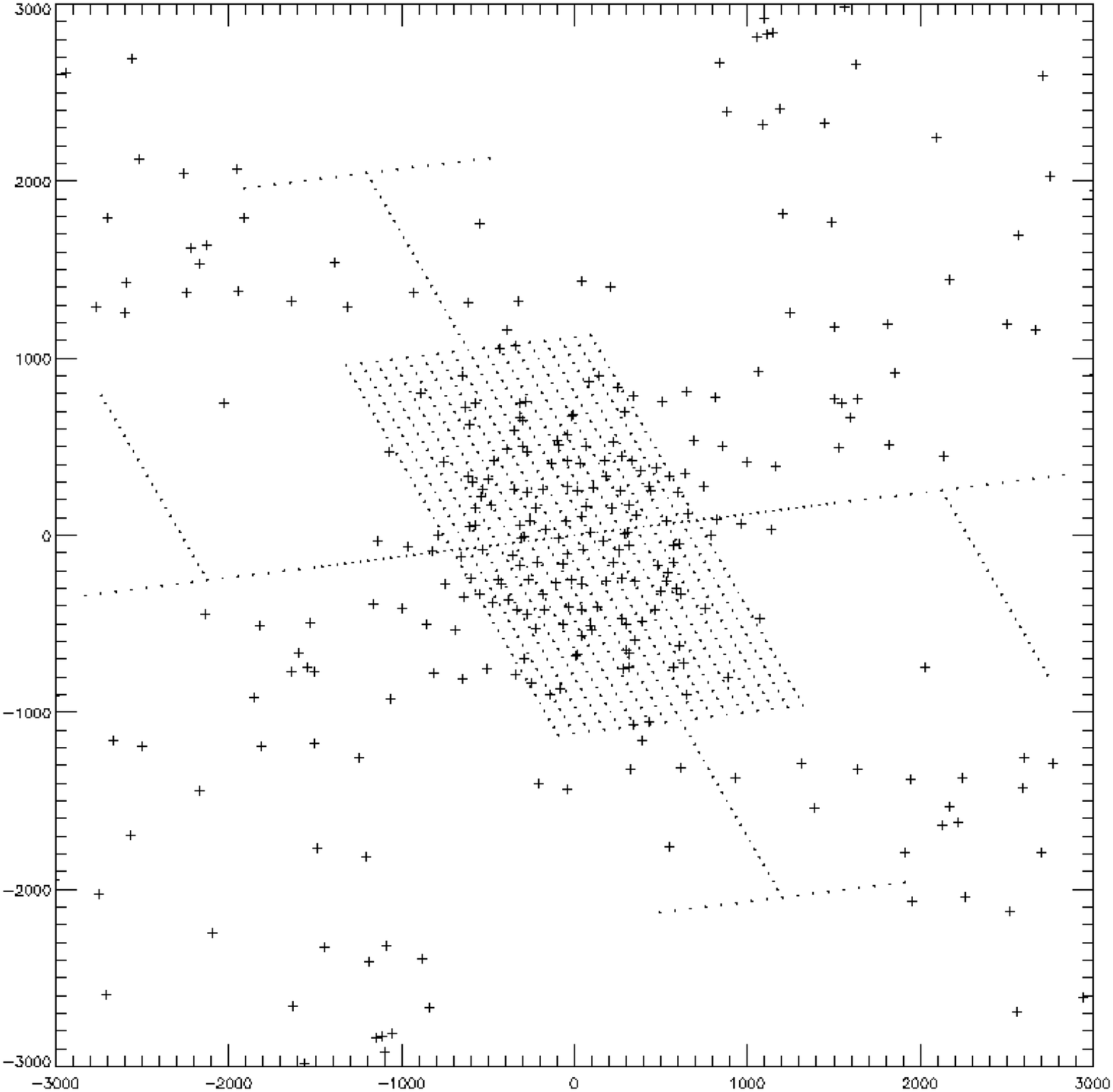}}
\caption{Actual $uv$ coverage ($rad^{-1}$) for 2002 Aug 27, 09:30 $UT$ with 
$NRH$ + $GMRT$. a) coverage with baselines up to 12 000 $\lambda$, b) zoom-in 
of the central area showing the overlap between $NRH$ and $GMRT$ coverages. 
$GMRT$ data are marked by $+$, while $NRH$ data are marked by dots.}
\end{figure}

\begin {enumerate}
\item		
The $NRH$ produces the central regular pattern (see fig. 1b): a densely 
covered parallellogram with some linear extensions, limited to values of
$r_{uv}=\sqrt{u^{2}+v^{2}}$ smaller than $\approx 3000 \lambda$. This results 
from the regular antenna distribution of the $NRH$.
\item		
The $GMRT$ produces an irregular and much more extended pattern, consisting of:
\begin {itemize}		
\item
a central set of points corresponding to baselines from the central compact 
array of $GMRT$, which has a substantial overlap with the parallellogram
of the $NRH$. We refer to the area of overlap between the inner baselines 
of the $GMRT$ and $NRH$ as $A_{\rm overlap}$.
\item
similar patterns with fewer points, distributed over 6 irregular arms,which 
correspond to baselines between the central array and the arm antennas.
\item
isolated points corresponding to baselines between the arm antennas.
\end {itemize}			
\end {enumerate}		
The composite dirty lobe $L_{d}$ is the $FT$ of this $uv$ coverage and is the 
sum of $NRH$ and $GMRT$ dirty lobes $L_{d \,RH}$ and $L_{d \,GMRT}$. Due to 
the regular $NRH$ uv coverage $L_{d \,RH}$ comprises of a pencil beam with 
classical ``sinc'' side lobes repeated over a period $\approx 1^{o}.$ (larger 
than the size of the sun) on both axes (Fig. 2a). 

\begin{figure}
\centering
\subfigure[]{
\includegraphics[width=3.85in]{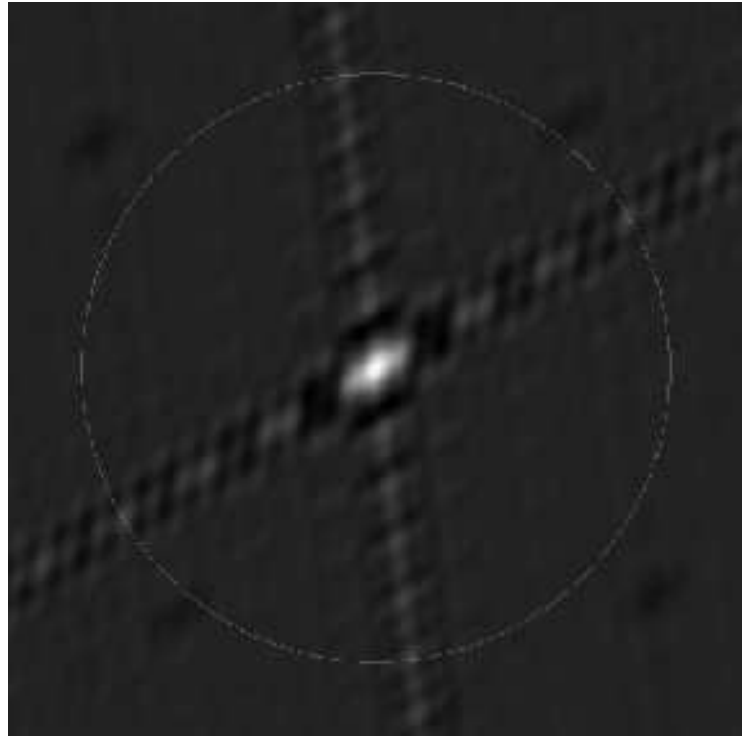}}
\hspace{1in}
\subfigure[]{
\includegraphics[width=3.85in]{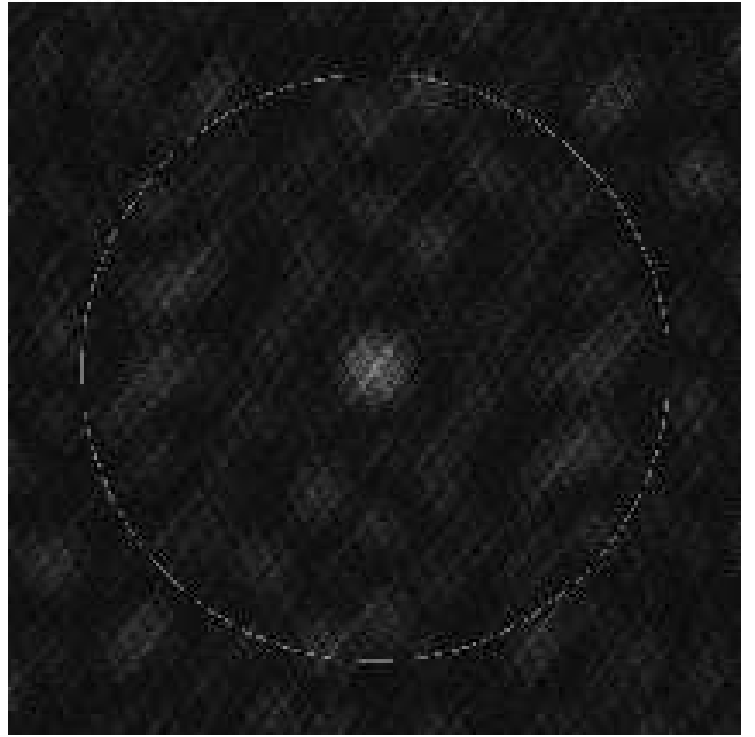}}
\caption{Theoretical lobes for (a) $NRH$ and (b) $GMRT$. The circle 
represents the solar optical limb.}
\end{figure}

Within one period, most 
of the flux of $L_{d \,RH}$ is concentrated in a small area surrounding the 
main pencil beam, with extent $\approx$ 2 or 3 times the width of the pencil. 
The resulting dirty image of the sun looks like a smoothed version of the sun,
with side lobes $<$ 20\% near compact sources.

The situation is somewhat different for $L_{d \,GMRT}$. The $GMRT$ $uv$ 
coverage is much more sparse and irregular and the beam is very complex (see 
fig. 2b) : the narrow main pencil beam contains only a small fraction of the 
flux and is sourrounded by a complicated pattern of large side lobes. At 
larger distances from the main pencil beam, the structure of lobe is aperiodic 
(no regular grating lobes as for the $NRH$) and exhibits features of various 
scales and amplitudes, the most intense ones being at distances somewhat 
smaller than the size of the sun. As a result, the $GMRT$ alone : 
\begin {itemize}		
\item
cannot give a reliable snapshot image of the whole sun, especially of the 
large scale features,
\item
the dirty image of several sources on the sun is complex and difficult to
understand visually, and requires a cleaning procedure.
\end {itemize}			
Similarly, the benefits of merging $GMRT$ data with $NRH$ data is not clear 
from a visual inspection of the dirty composite images (which appear visually
to be similar to $NRH$ dirty images plus some ``small scale noise''). The 
benefits are obvious only from the composite clean images.  


\section {Data selection and processing}
We present observations of the sun at 327 MHz on Aug 27 2002 between 9 and 
10 $UT$ in this paper. We chose this day because there are complex sources on 
the sun, which present a challenge for snapshot imaging. The $GMRT$ 
observations were integrated for a duration of 17 seconds, while the $NRH$ 
data were integrated for a duration of 0.125 seconds.

\subsection {Pre-processing}
The $GMRT$ data were separately pre-processed in the following manner, using 
the standard $AIPS$ software: a suitably interference free channel was first 
identified from the 128 channels spanning 16 MHz around 327 MHz. The data were
then rigorously edited for bad antennas and baselines. The visibilities on the
sun were then calibrated using those from the nearby phase calibrator 1021+219.
Before any subsequent processing, the $NRH$ data were therefore integrated 
over the 17 second intervals defined by the $GMRT$ data. The high time 
resolution $NRH$ data were first integrated over the 17 second intervals 
defined by the $GMRT$ data, in order to match the time cadence of the $GMRT$ 
data. The $NRH$ and $GMRT$ visibilities were then intercalibrated in the 
following manner: 

\subsubsection {Amplitude intercalibration}
The visibilty amplitudes from $NRH$ and $GMRT$ should be 
coherent, since both instruments use cosmic calibrators. $NRH$ and $GMRT$ $uv$ 
coverages overlap particularly densely on the area $A_{\rm overlap}$ around 
the origin of the $uv$ plane (corresponding to baselines $<1km$). Hence the 
sets of samples of the visibility on this area from both instruments should 
have comparable amplitude and merge homogeneously. This is especially true for
the sun, for it dominates the system temperature for both the individual 
instruments. We noted that the amplitudes of the $GMRT$ visibilities were 
larger (as compared to the NRH ones) by roughly one order of magnitude. This 
is primarily because the GMRT antennas are not fitted with solar noise 
calibrators, and the so-called ``$T_{\rm sys}$'' correction (Bastian 1989) 
can therefore not be applied. In other words, we cannot be certain about the
absolute values of amplitudes of the $GMRT$ visibilities (when observing the 
sun). We applied a multiplicative factor $a_{\rm intercal}$ to the $GMRT$ 
visibilities. The value of $a_{\rm intercal}$ was determined in such a way
that the mean value for the modulus of visibility, when averaged over 
$A_{\rm overlap}$, was the same for both sets of data. We estimate that the 
accuracy of the resulting intercalibration is a few \%. Visual inspection of 
amplitude versus $uv$ distance plots for the composite data convinced us that 
this simple procedure gives satisfactory results.

\subsubsection {Position intercalibration} 
The absolute positions of sources as observed by the two instruments can 
differ for several reasons: 
\begin {enumerate}
\item 
For merging the visibilities, data are extracted from the standard processing 
algorithms of each instrument. Since fringe stopping for solar observations is
complex and not necessarily done in one step, this can produce slowly varying 
position differences between $NRH$ and $GMRT$ data at this stage.
\item 
The inhomogeneities due to gravity waves in ionospheric electron density can 
produce apparent shifts of the order of 1$^{'}$ at $327$ MHz, varying on 
time scales $\sim30$ min (see for instance Mercier, 1996 and references 
therein). These effects are of local origin and would be totally uncorrelated 
between $NRH$ and $GMRT$ sites.
\end {enumerate}		
It is obvious that the position intercalibration must be better than the 
expected resolution. This was achieved in the following manner: for each 17 
second interval, $2D$ dirty images were separately calcultated for $NRH$ and 
$GMRT$ through a fast Fourier Transform (fft), including only baselines in the 
range common to both instruments (baselines $<$ 2 $km$, which makes this range 
somewhat larger than $A_{\rm overlap}$), in order to achieve comparable 
resolutions. We then defined the translation required to make the maximum of 
$GMRT$ dirty image coincide with the maximum of $NRH$ dirty image. This 
enables us to calulate the phase shifts that must be applied to all $GMRT$ 
baselines in order to make the spatial origin identical for both instruments.
We find that these differences in position were $\approx$ 0.15 $R_{S}$ and
vary by less than 0.02 $R_{S}$ for the duration of our observations (9--10 
$UT$, Aug 27 2002).

\subsection {Processing}
The processing comprises essentially of two steps : getting a ``dirty'' image
and cleaning it, in order to make it visually understandable. When a single 
intense compact source is present, we can also estimate its dimensions 
by fitting a 2-D Gaussian directly to the observed visibilities. As we will 
see below, this allows us to measure sizes smaller than what would have been 
possible via inspection of clean composite images.
\subsubsection {Dirty image}
The dirty image is merely the Fourier Transform ($FT$) of the joint set of 
visibilities. The fft algorithm has the advantage of being efficient, but it 
requires that spatial frequencies corresponding to the various baselines are 
distributed on a regular square grid in the $uv$ plane. This is not the case 
because $NRH$ and $GMRT$ baselines are not simply commensurable ($GMRT$ 
antennas are randomly distributed). This situation is usual in radioastronomy 
and regridding is commonly used. It consists of replacing the observed samples
of the visibility by samples interpolated on a regular grid. We show hereafter
that it is not convenient for snapshot images of a complex and wide object 
such as the sun with $NRH$ and $GMRT$.

According to Shannon's theorem, if $w$ is the total angular width of the sun, 
its visibilty must be sampled with a step $u_{sample}$ which satisfies :
\begin{equation}
u_{sample}<\frac{1}{w}		
\end{equation}
Taking $w\sim 45^{'}=$ 0.013 $rad$ at 327 MHz, $u_{sample}=77 rad^{-1}$. The 
corresponding distance between neighboring antennas (projected normal to the
line of sight) is $\lambda \times u_{sample} = 69m$. The sampling condition 
(1) is fullfilled for $NRH$ baselines (at least far from local noon for 
observations before January 2003) but not for the $GMRT$ baselines, especially
for large baselines involving the distant arm antennas (which yield the 
enhanced spatial resolution). Thus interpolated values of the visibility for 
large baselines can be strongly biased and produce spurious effects such as 
``ghosts'' of the actual sources. We emphasize that the condition (1) should 
be strictly satisfied for interpolating the visibility, since there can be 
several sources on the sun that are separated by more than 2 $R_{S}$.

In order to overcome this difficulty, we have defined a grid in the $uv$ plane
with a period $u_{grid} << u_{sample}$. In this way the observed visibilities 
can be ascribed to the nearest node of the grid, with only a small error. In 
other words, we keep the exact values measured for the visibility and make a 
small error (``round-up'' error) $< 0.5 u_{grid}$ on the spatial frequencies, 
which is much smaller than the minimum possible scale of variation of the 
visibility in the $uv$ plane, according to (1). In this way, we avoid the 
problem of interpolation in the $uv$ plane, at the expense of handling large 
arrays in fft computations.

In our data processing we used arrays of dimension $1536 \times 1536$ and 
restricted the longest baseline to $17 000 \lambda$ (beyond this value $GMRT$ 
$uv$ coverage is too sparse for snapshots, see \S~3.2.2 below). Hence 
$u_{grid} \approx 22 rad^{-1}$ and the round-up error in spatial frequencies 
is $ <11 rad^{-1}$, which is $<< u_{sample}$. Since $NRH$ and $GMRT$ taken 
together usually give fewer than 1000 visibility samples, most of the elements
of the 1536 $\times$ 1536 array in the $uv$ plane are zero. When computing the
image through a fft, the resulting field is $1/u_{grid} \approx 8 \,deg.$. 
This field is much larger than the maximum width of the sun observable with 
the $NRH$ (see eq.(1)), and we therefore retain only its central part.

\subsubsection {Cleaning procedure}
We first briefly recall the basic idea of $CLEAN$ deconvolution. When the $uv$
coverage is sparse and irregular, the dirty lobe $L_{d}$ has a complicated 
shape, with large side lobes at distances from the main peak that are much 
larger than its half power beam width ($HPBW$). The contribution of a point on
the object is not localized within one $HPBW$ of $L_{d}$ on the dirty image, 
and this makes the dirty image difficult to interpret visually. $CLEAN$ 
replaces the convolution of each point with the dirty lobe by a convolution 
with a clean lobe $L_{c}$ (which is usually a Gaussian, with no sidelobes). The
amplitude and $HPBW$ of the clean lobe are similar to that of the dirty lobe. 
The procedure is iterative, starting with the brightest points on the dirty 
image. The clean image thus depends ``locally'' (i.e., within one $HPBW$) on 
the object and is thus visually more understandable. For technical reasons the
replacement is only partial at each step (see for instance Thompson et al, 
1986, chapter 11). In the $uv$ domain, the $CLEAN$ process can be viewed as 
one where the holes in the $uv$ coverage are filled up via interpolation. 
However it is well known that $CLEAN$ works rather poorly for continuous 
objects that have a wide spectrum of spatial scales. Some remedies have been 
proposed to alleviate this problem. In particular, when the object involves 
two well separated spatial scales (eg., few compact sources superposed on the 
quiet sun), a two-scale $CLEAN$ procedure can be used (Wakker and Schwarz 
1991). One can also try to subtract a model of the quiet sun from the dirty 
image. $CLEAN$ then works on a dirty image which supposedly results only from 
only some isolated compact sources. The model of quiet sun is added afterwards 
to the cleaned image for obtaining a final clean image.

The cleaning procedure we developed here is a multi-scale version of $CLEAN$ 
($MS-CLEAN$) adapted to objects with features of various scales (Delouis 1998).
The basic idea is to first produce several filtered dirty images 
$I_{d 0}$, ... $I_{d n}$ from the original dirty image $I$. Each filtered 
dirty image involves a relatively narrow range of spatial scales. For this 
purpose, we use successive filters $F_{0}$, ..., $F_{n}$ in the $uv$ plane, 
each filter selecting progressively smaller spatial scales: 
 \begin {itemize}			
\item 					
$F_{0}$ is a low pass filter with cut-off $u_{e0} \approx$ 3--4 
$u_{sample}$ (recall that $u_{sample} \approx 3.5$ $u_{grid}$ and that the 
observable field of view is 1 / $u_{sample}$).
\item					
The filters $F_{1}$, ... $F_{n}$ filters are band pass filters with internal 
radii $u_{i1} = u_{e0}$, ... $u_{in} = a^{n-1} u_{e0}$ and external radii 
$u_{e1} = a u_{i1} = a u_{e0}$, ... $u_{en}=a^{n} u_{1}$, where $a \approx 3$ 
in order to limit the spatial scale ranges.
\end {itemize}				
The edges of these filters are smooth and their sum is unity.

The first filtered dirty image $I_{d 0}$ is very smooth and comprises of only 
a few independent pixels. It can be cleaned using smooth dirty and clean beams
$L_{d 0}$ and $L_{c 0}$ deduced from $L_{d}$ and $L_{c}$ through a filtering 
with $F_{0}$. Thus the largest structures in the dirty image can be described 
in terms of only a small number of independent smooth components. The problem 
of describing a large scale structure with a large number of narrow compoments 
(if the $CLEAN$ algorithm were directly applied) is then reduced to that of 
describing a source in components $\approx$ 4 times narrower than the widest 
source that can be described. It is obvious that the number of iterations 
(and with it, the possibility of producing artefacts) in the $CLEAN$ procedure
will be drastically reduced.

The same argument can be applied to other filtered dirty images : each case
is not very different from the case of a relatively small number of compact
sources (``compact'' relative to the dirty lobe). The final clean image
is simply the sum of the filtered clean images.

After some trials we found that the best results were obtained under the
following conditions:
\begin {itemize}	
\item 
The $GMRT$ $uv$ coverage gets increasingly sparse for progressively longer 
baselines. At 327 MHz there are only isolated points for baselines $> 17000 
\lambda$. This could lead to artefacts when cleaning dirty images from complex
objects. We found that the best compromise between reducing these artefacts 
and preserving the spatial resolution could be achieved by taking a clean lobe 
$L_{c}$ for which the width of its $FT$ is  $\approx 0.35 \times uv_{max} = 
0.35 \times b_{max} /\lambda$,  where $b_{max}$ is the longest baseline under 
consideration. In simulations we could take $uv_{max} =$ 17000 or 12000, 
whereas for real data we had to take $uv_{max} = 12000$ only, probably 
because of imperfect calibration. In other words, $MS-CLEAN$ works poorly
for sparse $uv$ coverages. The resulting $HPBW$ of the clean lobe was 
$\approx 1^{'}$.
\item
We took a maximum index $n$=3 for the filters (i.e., a central zone and 3 
concentric rings). This allows us to cover a spatial frequency range up to 
$\approx 10^{4} rad^{-1}$ with a ratio $a\approx 3$ for the ratio between the 
external and internal radii of filters.
\end {itemize}		

\subsubsection {Estimating the width of an intense compact source}
The resolution corresponding to the width of our final clean beam is much
lower than what could be expected from the $GMRT$ size. As explained earlier,
this limitation arises because we try to obtain images from a complex object 
with a coverage that becomes very sparse for long baselines. However, for 
particularly simple objects, (e.g., when the composite image contains a single 
intense, compact source) it is possible to estimate the source size by fitting
a gaussian model directly to the amplitude of the visibility over the whole 
$uv$ range observed by the $GMRT$. It is then possible to compare this derived 
size with what is directly measured on clean images. Of course, this method is 
meaningful only when there is a dominant compact source, and cannot be applied 
when several sources with similar intensities co-exist. 

\section {Results}
We first checked the expected performance of our method using simulated 
brightness distributions that are similar to the actual distribution on the 
day of our joint observations. We found that this procedure gives us a 
reliable method of anticipating the limitations when we process actual data. 
We show here the results of two representative simulations.

\subsection {Simulations}
\subsubsection {The models for the radio sun}
We defined a class of solar models which contain several features commonly 
observed at 327 MHz. They comprise structures of various sizes, ranging from 
the size of the quiet sun, down to 0.05 $R_{S}=52^{''}$. These structures are
derived from a simple mathematical model. The brightness distribution in the 
$1D$ analog of this model can be written as
{\begin{eqnarray}
\nonumber 
b(x) = \biggl(1-\biggl(\frac{x}{a}\biggr)^{e_{1}}\biggr)^{e_{2}} \, , 
\,\,\,\, -a<x<a\,, \\
b(x) = 0 \, , \,\,\,\, \mid x \mid > a \, . 
\end{eqnarray}}
The parameters a, $e_{1}$ and $e_{2}$ define the half-power width of the 
structure, the flatness of its maximum and the slopes of its sides, 
respectively. In practice, we took a $2D$ model with an elliptical section, 
where the tilt of the ellipse major axis is also a free parameter. The model 
of the sun is the sum of $\approx 30$ such structures with various positions, 
sizes, flatness, tilt and amplitudes. The structures are:
\begin {itemize}		
\item
a relatively flat structure for the quiet sun, with total extent 42 $\times
38^{'}$ ($EW\times NS$), and $T_{B}$ = $ 10^{6}K$.
\item
several ``thermal'' regions with various slopes, widths in the range 6 - 
$9^{'}$, various tilts and excess brightness of some $10^{5}K$ above the quiet
sun.
\item
two coronal holes, one very elongated, with steep sides and depression depths 
$\approx 3 \times 10^{5}K$,
\item
a number of compact elliptical ``non thermal'' sources with sizes 1 -- $3^{'}$,
various tilts and excess brightness of some $10^{5}K$--$3 \times 10^{7}K$ 
above the quiet sun,
\item
a faint CME-like arch whose brightness decreases with height and with distance 
from the central axis.
\end {itemize}			

\begin{figure}
\centering
\includegraphics[width=2.6in]{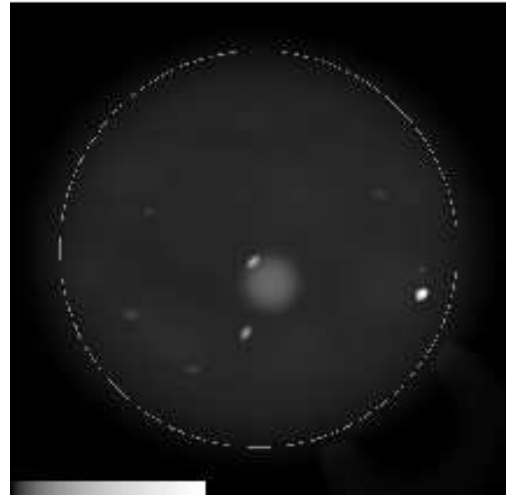}
\caption{Model 1 adopted for the simulations.}
\end{figure}

Using these ingredients, we produced two particular models for our 
simulations :
\begin {enumerate}
\item
A relatively simple model (model 1) in which only 4 ``non thermal'' sources
are intense ($T$ up to $3 \times 10^{7}K$) and therefore dominate the other
structures,
\item
A more complex model (model 2), which has ``non thermal'' sources with 
$T \sim 10^{5}K$ -- $5 \times 10^{6}K$, so that they do not dominate the other 
features as in model 1.
\end {enumerate}
\subsubsection {Model 1}		
Fig. 3 shows model 1. It is similar to several instances encountered in the 
real data of Aug 27 2002. It consists of :

\begin {itemize}			
\item a bright western source with $T_{B} = 3 \times 10^{7}$K,
\item two relatively dimmer sources near the disk center and  
extended emission (of size 4$^{'}$) between these two central sources.
\end {itemize}				
[The color scheme for fig. 3 (and the subsequent figures) was chosen so as to
optimize its visual appearance on paper prints. The intensity scale is 
indicated on the lower left corner: it is linear for low levels and 
progressively saturates at higher levels. This highlights low level features 
($CLEAN$ artefacts, in particular)].

\begin{figure}	    
\centering
\subfigure[]{
\includegraphics[width=2.6in]{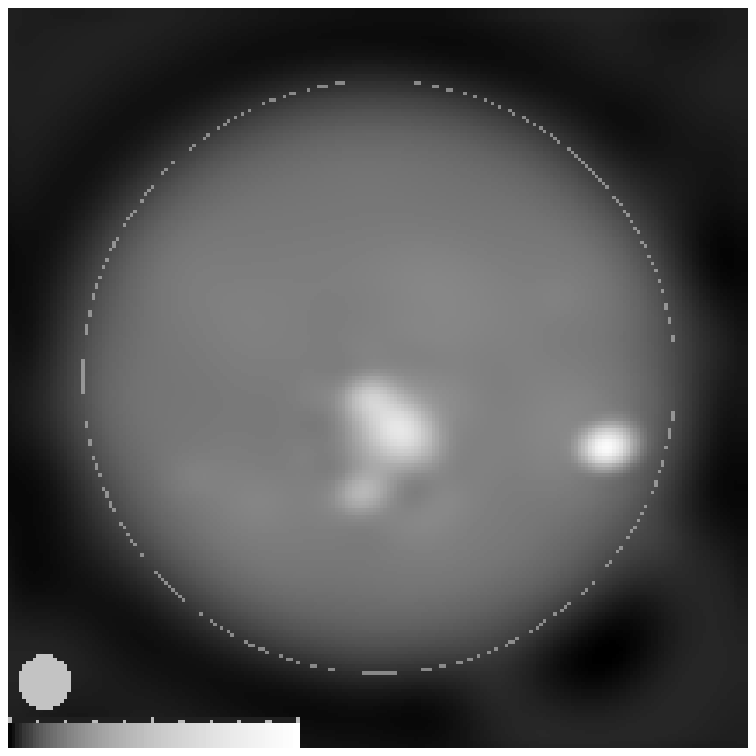}}
\subfigure[]{
\includegraphics[width=2.6in]{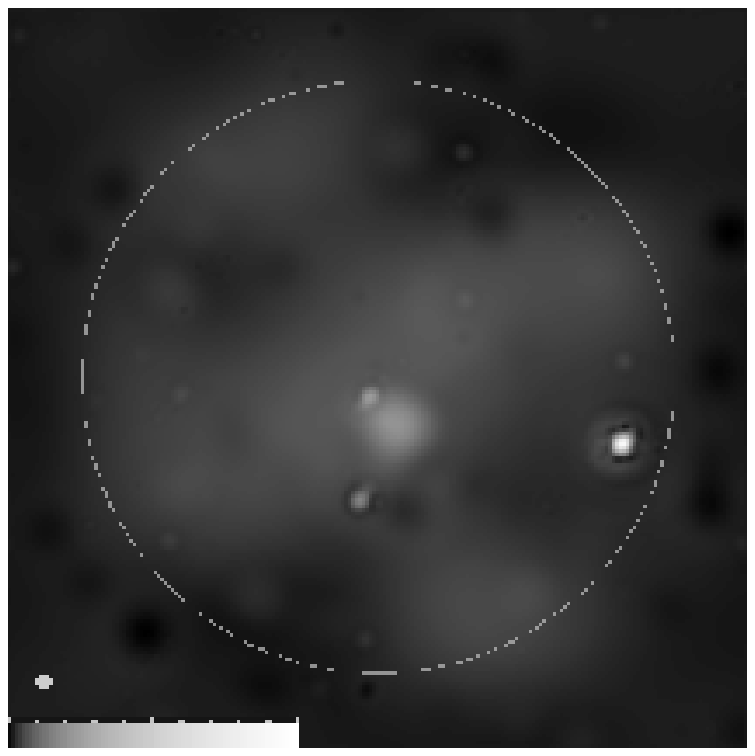}}
\subfigure[]{
\includegraphics[width=2.6in]{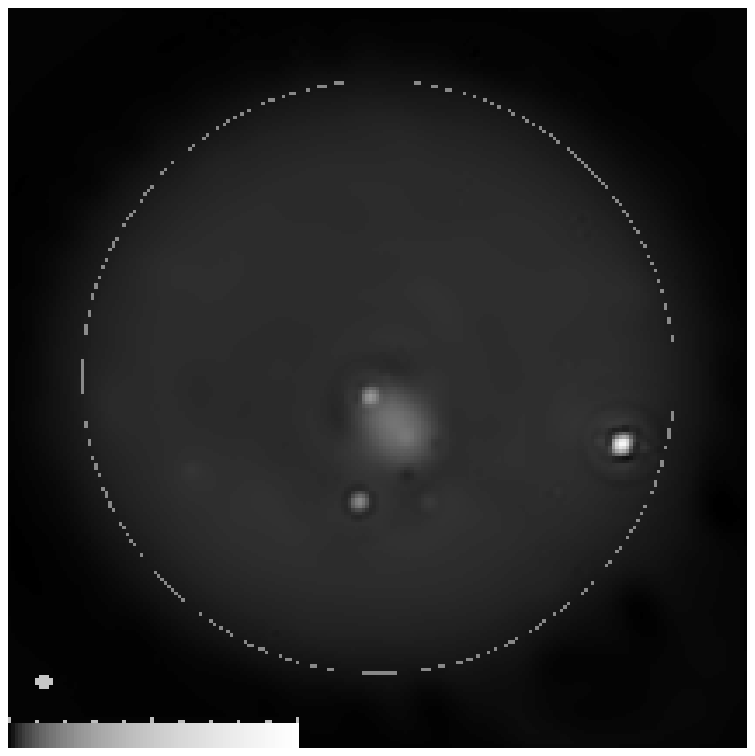}}
\caption{Clean image for model 1 : a) $NRH$ alone, b) $GMRT$ alone, c) 
$NRH$+$GMRT$. The circle indicates the optical solar limb. The size of the 
clean beam is indicated in the lower left hand corner.}
\end{figure}
All other features (including the quiet sun) are hardly visible. Fig. 4 shows 
the clean images obtained by the $NRH$ alone (a), $GMRT$ alone (b) and $NRH$ 
+ $GMRT$ (c) when using the actual $uv$ coverages provided by the $NRH$ and 
$GMRT$ at 09:30 $UT$ on Aug 27 2002. This means that the GMRT baselines which 
were flagged for the actual data of Aug 27 2002 were not used here either. The 
synthesized clean beam is shown at the bottom left corner of each figure. For 
cases b and c, it has an extent of 49$^{''}$.
Even for this relatively simple model, the improvement in quality of the 
composite clean image (c) is quite evident : the resolution is much better 
than the $NRH$-only image, and there are fewer spurious large scale structures
than in the $GMRT$-only image.

\subsubsection {Model 2}
Model 2 is shown in Fig. 5. 
\begin{figure}
\centering
\subfigure[]{
\includegraphics[width=2.6in]{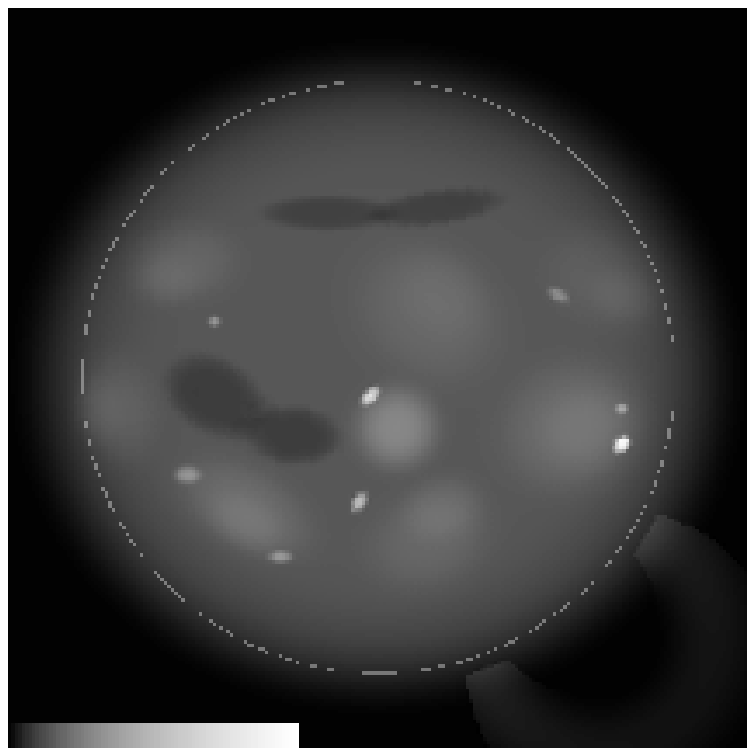}}
\subfigure[]{
\includegraphics[width=2.6in]{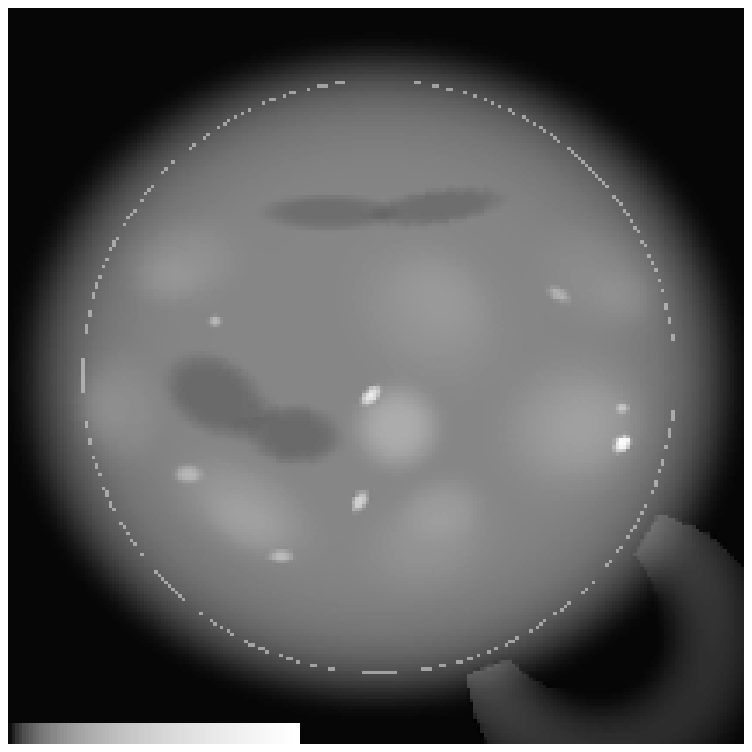}}
\caption{a) and b): Model 2 adopted for the simulations: a more complex sun. 
Panel a) shows an overview of the model.
The color scheme has been chosen to enhance lower levels in panel b).}
\end{figure}
It differs from model 1 in that the the nonthermal sources are dimmer, which 
allows us to discern the other features better: the quiet sun, patches of 
``thermal'' emission, coronal holes, faint compact sources and a CME-like 
arch in the SW. 

Fig. 6 shows the clean images obtained by the $NRH$ alone (a),
$GMRT$ alone (b) and $NRH$ + $GMRT$ (c), using the same $uv$ coverage 
as for model 1 : 
 
\begin{figure}  
\centering
\subfigure[]{
\includegraphics[width=2.6in]{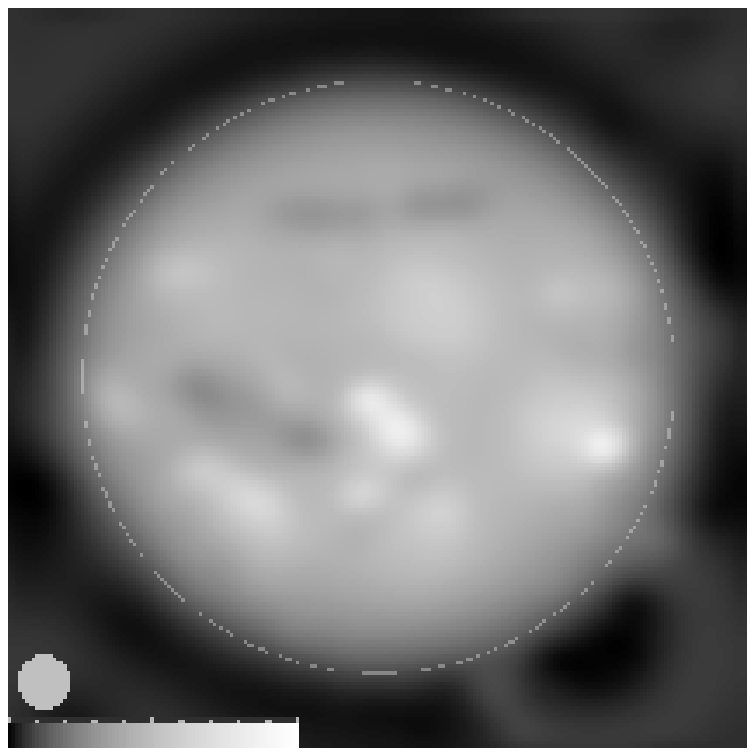}}
\subfigure[]{
\includegraphics[width=2.6in]{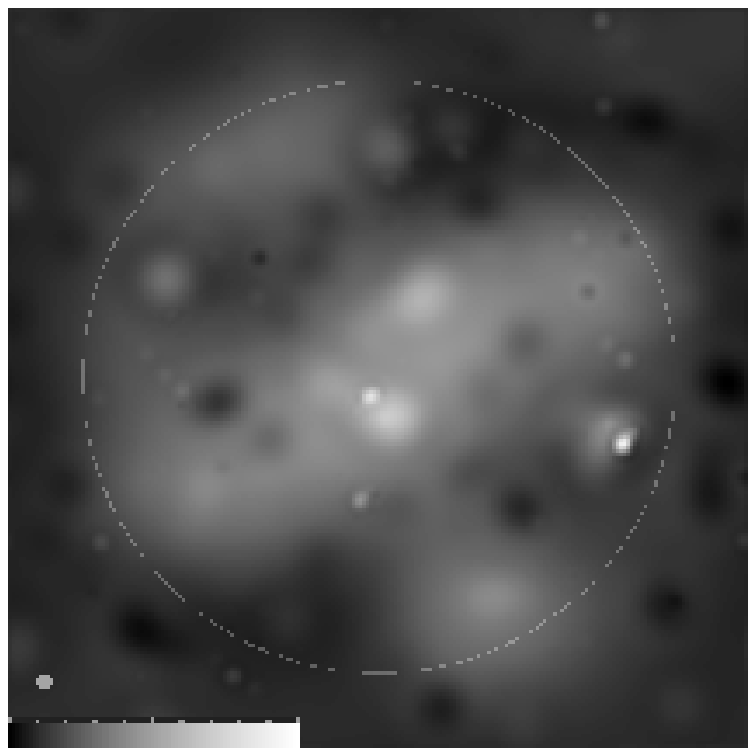}}
\subfigure[]{
\includegraphics[width=2.6in]{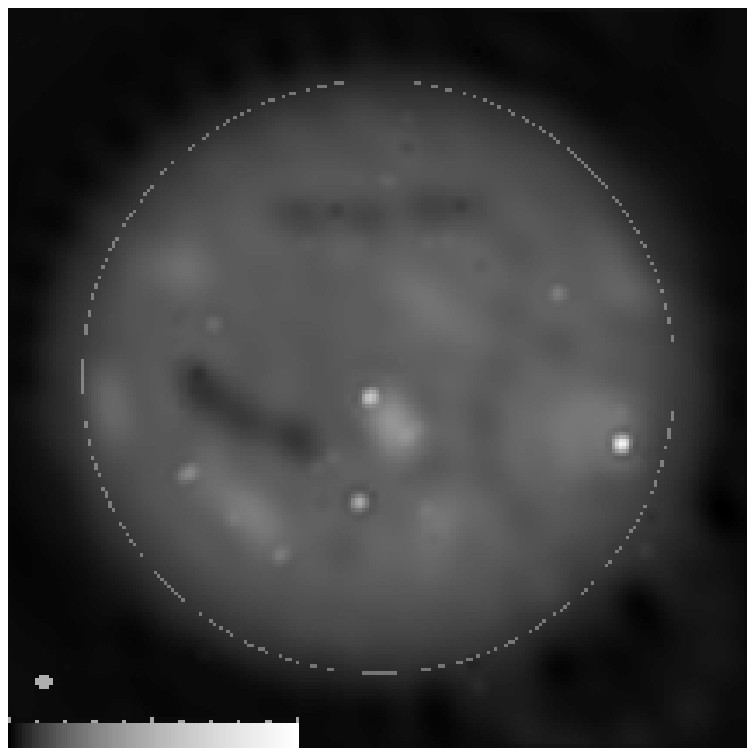}}
\caption{Clean image for model 2 : a) $NRH$ alone, b) $GMRT$ alone, c) $NRH$ 
+ $GMRT$.}
\end{figure}
The $NRH$ image is smooth (faint compact sources are not seen), but large 
scale structures are reliable, although they appear with a reduced contrast. 
The faint CME beyond the south west limb is satisfactorily evident.
Conversely, due to the greater complexity of the object, the clean image 
from $GMRT$ alone is practically unusable : large scale structures cannot be 
recognized, and only the two most intense compact sources can be identified. 
The improvement in quality of the composite image (c) is also evident, 
although limitations due to the sparse $uv$ coverage more apparent than for 
model 1 :
\begin {itemize}			
\item the edges of coronal holes are blurred
\item low level artefacts (both positive and negative) appear around coronal 
holes and thermal regions,
\item some of the faintest compact sources are difficult to distinguish from
artefacts.
\end {itemize}				

The CME is less visible on fig 6c than on fig 6a, but this is only due to the
higher contrast of fig. 6c, and because of its superior resolution.

Fig 7 shows the composite clean image obtained with the $uv$ coverage that 
could be provided if all $GMRT$ antennas were functioning. 
\begin{figure}  
\centering
\includegraphics[width=2.6in]{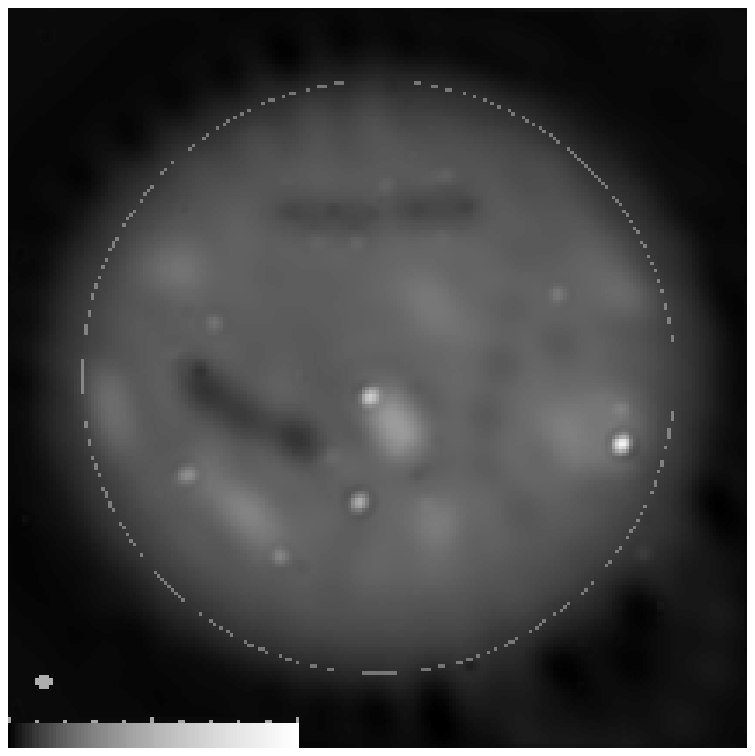}
\caption{Clean composite ($NRH$ + $GMRT$) image for model 2, using all the 
435 possible GMRT baselines}
\end{figure}
It is somewhat better than fig. 6c : faint compact sources are more reliable 
and the northern coronal hole has fewer artefacts. 

As expected, the benefits of merging $NRH$ and $GMRT$ data are obvious when
compact sources are present. However, it may be noted that low contrast 
extended features (e.g., the coronal holes of model 2) are also better imaged 
with the composite instrument. Model 2 illustrates the limits of what can be 
hoped from composite snapshots of a complex object with sparse extended $uv$ 
coverage.

In the preceding simulated examples, we had introduced no calibration errors
(i.e., deviations in gains and phases). We have also produced images including
random phase deviations in the simulated visibilities as a simple means of 
approximating the effect of phase calibration errors. We found that 
calibration errors are the principal cause of clean artefacts on the final
images, and can effectively limit the resolution : for $rms$ phase deviations
$>$ 20 $deg$, clean artefacts become noticeable and the extent of the 
accepted $uv$ coverage needs to be reduced. 

\subsection {Observed data}
We now turn our attention to the data actually observed between 09:02:23 $UT$
and 09:57:54 $UT$ on Aug 27 2002. During this time, there was a strong ongoing
noise storm near the west limb (0.69 $R_{S}$ west, 0.34 $R_{S}$ south), 
hereafter referred to as ``$W$''. There were also two weaker noise storm 
sources near the disk center (0.13 $R_{S}$ east, 0.12 $R_{S}$ south and 0.13 
$R_{S}$  east, 0.34 $R_{S}$ south), referred to as ``$N$'' and ``$S$'' 
respectively. While the western noise storm source remains compact throughout 
the duration of our combined observations, the two sources near disk center 
evolve significantly, showing marked extensions at certain times. We have 
selected three 17 second snapshots, which highlight some of these features and
illustrate the capability of our combined imaging technique.

Figure 8 shows 17 second snapshots centered around 09:20:25 $UT$, using data
from $NRH$ alone (a), $GMRT$ alone (b) and from $NRH$ + $GMRT$ (c).
\begin{figure}   
\centering
\subfigure[]{
\includegraphics[width=2.6in]{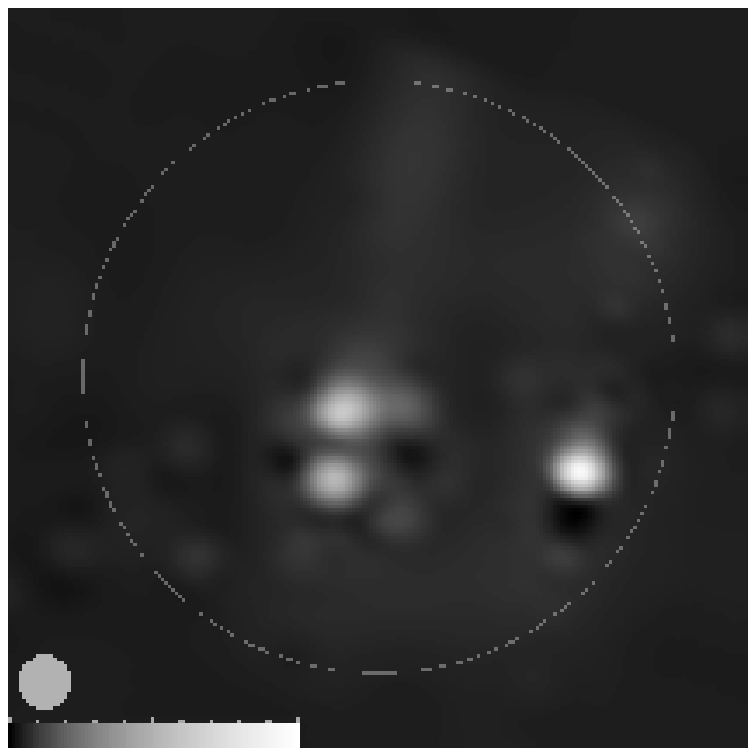}}
\subfigure[]{
\includegraphics[width=2.6in]{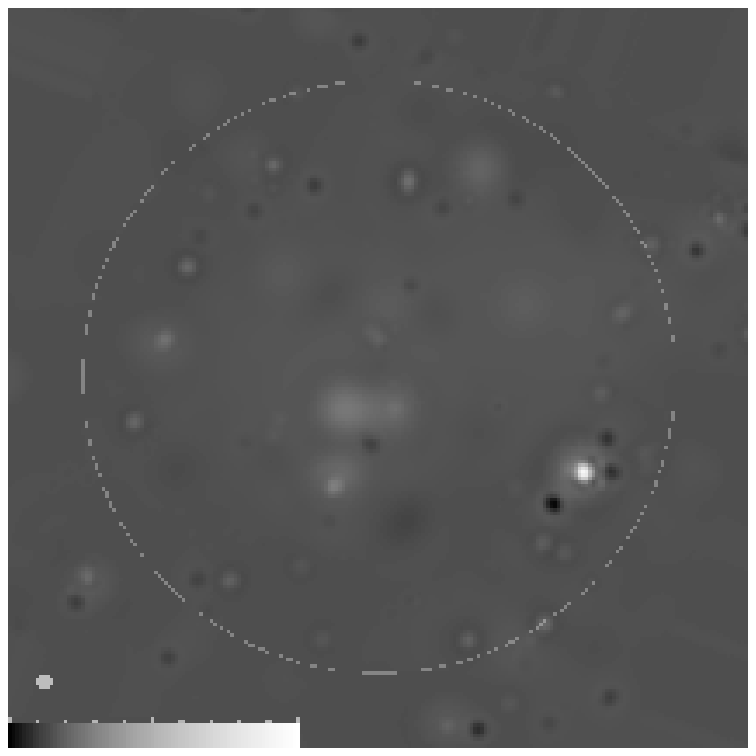}}
\subfigure[]{
\includegraphics[width=2.6in]{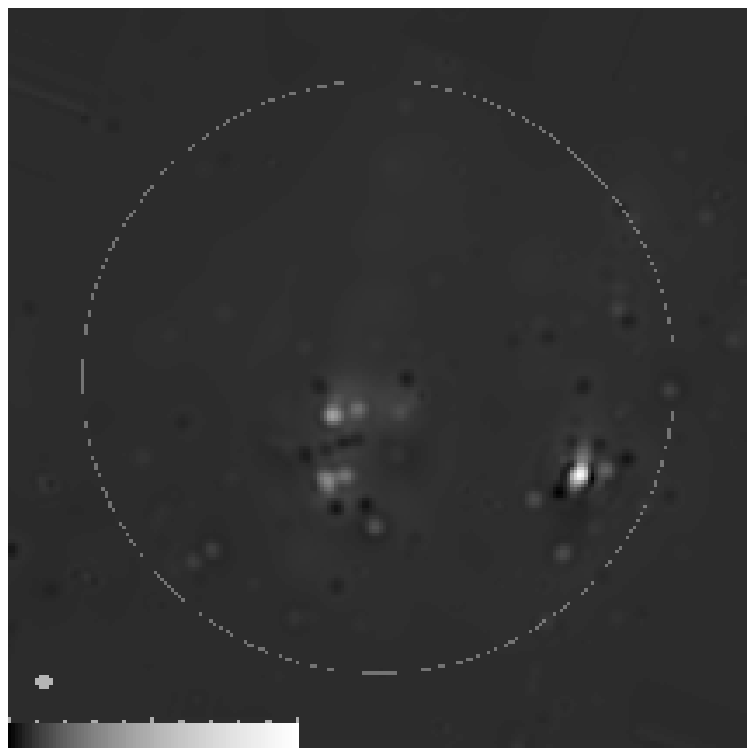}}
\caption{Clean image at 09:20:25 $UT$ from a) $NRH$ alone, b) $GMRT$ alone, 
c) $NRH$ + $GMRT$.}
\end{figure}
The $uv$ coverage has already been shown in fig. 1 : the relatively sparse, 
extended $uv$ coverage of the $GMRT$, which is complemented by the dense, 
short baseline $uv$ coverage of the $NRH$.

The improvement in resolution (when the GMRT data are used) is evident : the 
half-power size of the clean beam in fig. 8b and 8c (lower left hand corner) 
is 49$^{''}$. The $rms$ dynamic range for the composite image (fig 8c) is 244,
while the $max$ dynamic range is 11. On the $GMRT$-only image (fig 8b), we 
note that :
\begin {itemize}
\item artefacts are spread over the whole field. The relatively strong level
of negative artefacts produces a grey background for the picture.
\item the two central sources are not properly imaged. This is especially true
of ``$N$'' which is smooth in fig 8b, whereas it is more compact on the 
composite image (fig 8c).
\end {itemize}

The composite image shows artefacts which are both fainter (as is evident from
the darker background) and more localized to the vicinity of sources. This 
must be compared to simulations with model 1 (which is very similar to this 
actual case), for which $GMRT$ alone gives satisfactory results for compact 
sources and a much lower level for artefacts (fig. 4b). Since the data 
processing is the same for the simulations and for real data, we think that 
the difference should be ascribed to imperfect phase calibration of $GMRT$ 
data, and possibly to phase shifts arising from the 30 dB solar attenuators, 
which are switched on for solar observations and off for calibrator 
observations. 

The western source ``$W$'' is brightest at 09:13:39. Fig. 9 shows clean
snapshot images at this time from $NRH$ alone (a), $GMRT$ alone (b) and from 
$NRH$ + $GMRT$ (c). 
\begin{figure}   
\centering
\subfigure[]{
\includegraphics[width=2.6in]{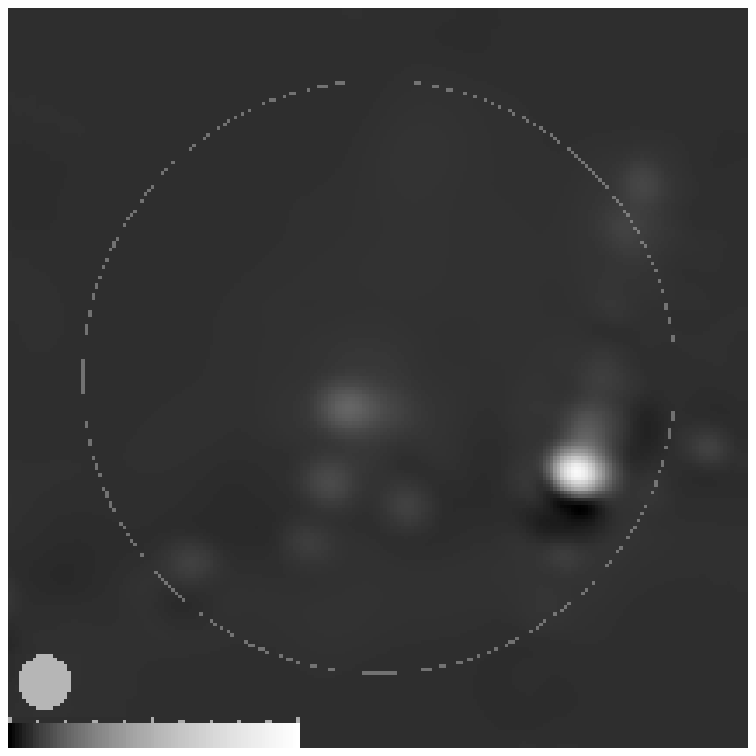}}
\subfigure[]{
\includegraphics[width=2.6in]{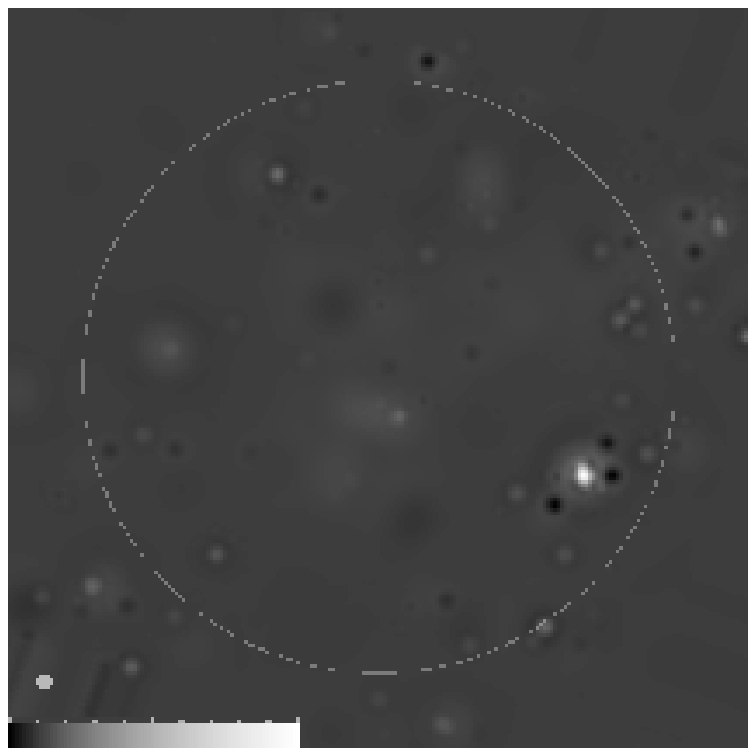}}
\subfigure[]{
\includegraphics[width=2.6in]{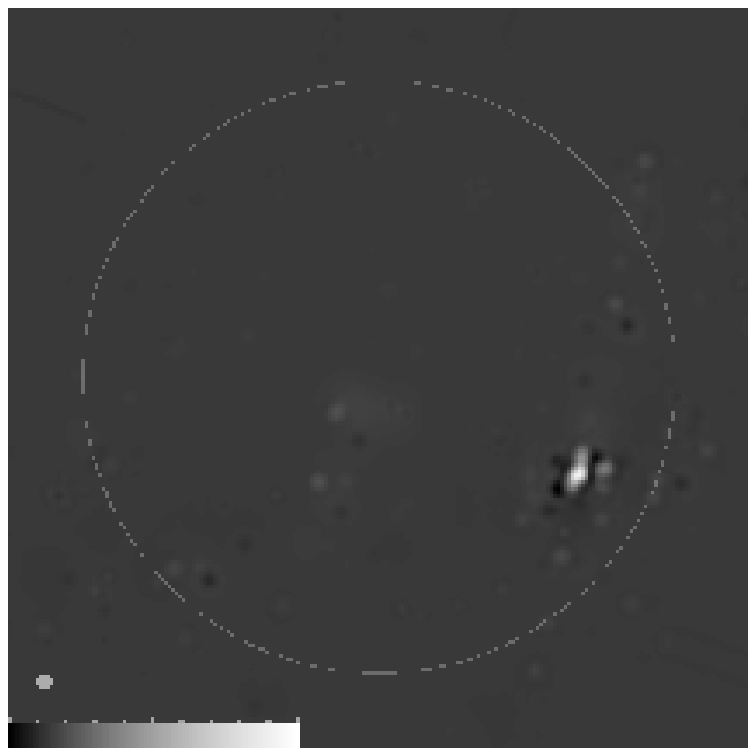}}
\caption{Clean image at 09:13:39 $UT$ from a) $NRH$ alone, b) $GMRT$ alone, 
c) $NRH$ + $GMRT$. Resolutions are the same as for Fig 8.}
\end{figure}
Clearly, the improved resolution is provided by the $GMRT$ baselines, while 
$NRH$ data helps in reducing clean artefacts. Note that, as for the preceding 
case, ``$N$'' is also poorly imaged by the $GMRT$ alone. The resolution of the
composite image, as in earlier cases, is 49$^{''}$. The $rms$ dynamic range of 
this image is 370 and the $max$ dynamic range is 21. 

The source ``$W$'' in this snapshot is the smallest structure present in
our observations. It is barely resolved with the clean beam size of 49$^{''}$.
However, since there is only one intense source present at this time, we can
use the step described in \S~3.2.3 (fitting a gaussian to the visibilities) to 
estimate a size of  $\approx 1^{'}$. Using simulations, we checked that this 
derived size is not appreciably affected by the faint sources ``$N$'' and 
``$S$'' and by the faint halo surrounding ``$N$''. High time resolution $NRH$ 
data show that the sources ``$N$`` and ``$S$'' (and ``$W$'', to a lesser 
extent) exhibit continual small-scale motions. Since the time resolution of 
composite images is limited by the 17 second integration time of the $GMRT$ 
data, it is possible that this snapshot image has ``smeared over'' such small 
scale motions. It follows that the source size of 1$^{'}$ we derive for 
``$W$'' is likely an over-estimate of its actual size.

Fig. 10 shows 17 second snapshots centered around 09:04:04 on Aug 27 2002,
from  $NRH$ alone (a), $GMRT$ alone (b) and from $NRH$ + $GMRT$ (c). 
\begin{figure}  
\centering
\subfigure[]{
\includegraphics[width=2.6in]{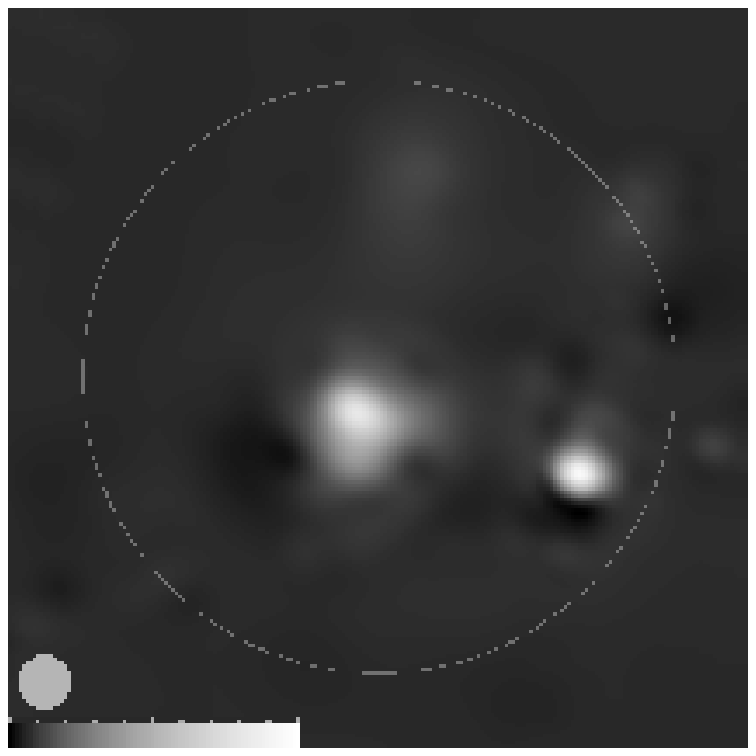}}
\subfigure[]{
\includegraphics[width=2.6in]{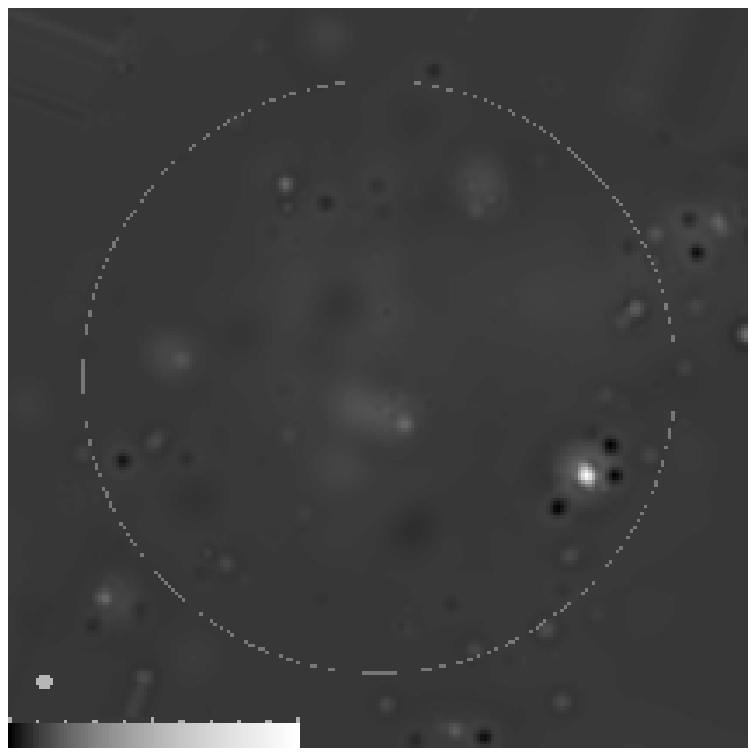}}
\subfigure[]{
\includegraphics[width=2.6in]{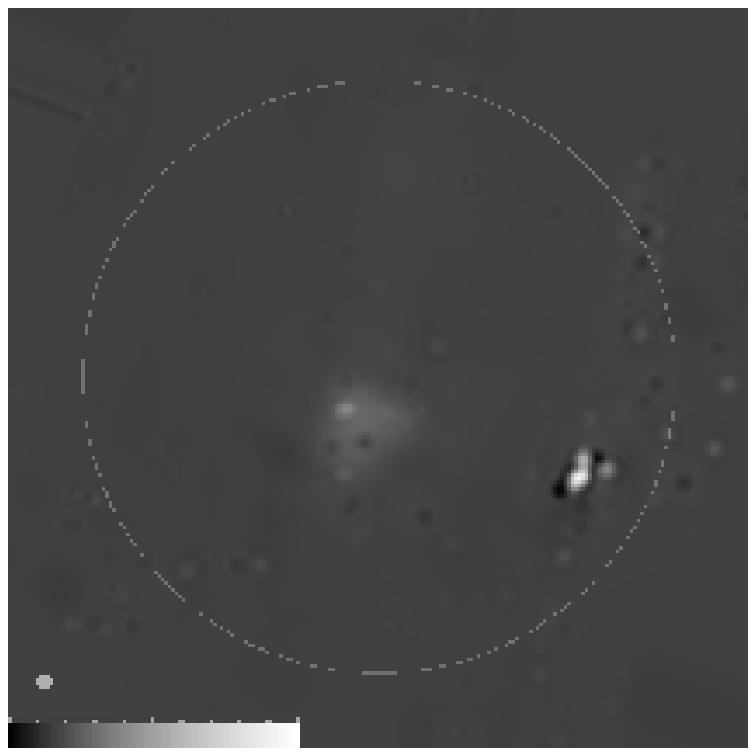}}
\caption{Clean image at 09:04:04 $UT$ from a) $NRH$ alone, b) $GMRT$ alone, 
c) $NRH$ + $GMRT$. Resolutions are the same as for Fig 8.}
\end{figure}
It illustrates the ability of the composite instrument to resolve and image 
extended emission reliably. 
The extended emission of around 200$^{''}$ that lies in between, and to the 
west ``$N$'' and ``$S$'' is imaged much better in the composite image than in 
any of the individual images. None of the instruments can give a satisfactory 
image of ``$N$'' : the resolution of $NRH$ is too low, and $GMRT$, in spite of 
its high resolution, does not image the bright core of ``$N$'' properly. The 
resolution of this composite image is 49$^{''}$ as for the other images. The 
$rms$ dynamic range of this image is 280, and the $max$ dynamic range is 14.

Upon an examination of high time resolution NRH data, it is evident that the 
relatively faint halo near ``$N$'' on the composite images is in fact a group 
of short, bright and almost unpolarized type III bursts, beside the noise 
storms ``$S$'' and ``$N$''. In addition to the short time scale motions of 
``$S$'' and ``$N$'' mentioned earlier, this shows that the observations 
presented here are only illustrative of the possibilities of the joint 
instrument. Higher time resolution observations with this joint instrument are 
very much needed in order to derive physical interpretations of solar radio 
emissions in this frequency range.

\section {Summary and conclusion}
We have demonstrated the capablities of a new $NRH$ + $GMRT$ composite 
instrument for obtaining snapshot images of the sun at 327 MHz. We have 
shown examples from simulations with two models (of differing complexity) 
of the radio sun, and from real observations of complex, evolving noise 
storm sources on Aug 27, 2002.

The simulations show the capabibility of the joint instrument, to produce
snapshot images of a complex sun, with structure sizes ranging the resolution
($\approx 50^{''}$) up to the size of the whole sun ($\approx 40^{'}$). The
main conclusions from the simulations are :
\begin {itemize}
\item
$GMRT$ and $NRH$ are highly complementary : long $GMRT$ baselines provide 
resolution, whereas dense $NRH$ $uv$ coverage at short baselines prevent 
aliasing of large structures which can occur with $GMRT$ alone.
\item
The composite instrument allows a substantial advance in snapshot imaging of 
complex objects. The resolution and dynamic ranges of composite images are far
better than those of images from the individual instruments (Fig 7).
\item
The quality of the calibration can limit the resolution and produce appreciable
artefacts.
\end {itemize} 		

We produced composite 17 sec snapshot images (from actual observations of the 
sun on Aug 27 2002) of structures between 60$^{''}$ and $\sim 200^{''}$ in 
size with a resolution of 49$^{''}$ and $rms$ dynamic ranges of 250--420. 
The quality of the composite image is far better than those of images from the
individual instruments (eg fig. 8). The $max$ dynamic ranges of the snapshot 
maps are $\gtrsim 20$. To the best of our knowledge, these are the 
highest dynamic range snapshot maps of the sun at meter wavelengths. Until 
now, high dynamic range radio maps of the sun were typically made by synthesis
imaging over time periods of a few hours. As mentioned in the introduction, 
high dynamic range images would be essential in studying phenomena like bright
radio bursts occurring along with (fainter phenomena like) coronal mass 
ejections. 

These observational results are only illustrative of the capabilities of the
technique we have employed. Merging $GMRT$ and $NRH$ data will potentially 
yield $\approx$ 1000 baselines per snapshot and would yield very high 
resolution, high dynamic range maps. This is roughly equivalent to
simultaneously using the capabilities of the VLA B, C and D arrays (the
comparison concerns only the number of baselines, since these VLA 
configurations cannot be used simultaneously to produce snapshots).

The resolution of the maps presented in this paper was limited to 49$^{''}$, 
which is well below what one would crudely expect from the $GMRT$. This 
limitation in the resolution is mostly because:
\begin{itemize}
\item
We are dealing with snapshot images and the $uv$ coverage of the $GMRT$ is
sparse, especially for large baselines. The deconvolution procedure does not 
work well for complex objects under these conditions. We need to introduce 
substantial tapering of the visibilities (see \S~3.2.2) in order to limit 
artefacts. This is a basic limitation on snapshot images, which would not 
concern synthesis composite images because the $uv$ coverage (for synthesis 
images) would be much less sparse, especially for large baselines. When the 
object is simple enough (eg fig. 9), however, one can achieve (from snapshot 
data) the full resolution corresponding to the size of $GMRT$ by using 
classical model fitting. 
\item
We had to flag several $GMRT$ baselines owing to malfunctioning antennas and 
radio frequency interference.
\item
We encountered calibration problems with $GMRT$ data. We have checked the
effect of calibration errors with simulations, and it is clear that they
make a significant difference in the level of clean artefacts in the final 
images. The resolution of 49$^{''}$ we have used is the best compromise 
between achieving the best possible resolution and maintaining a reasonably 
low level of clean artefacts.

\end{itemize}

For future observations, it is desirable to:
\begin{itemize}
\item
Achieve the best possible calibration, especially for the $GMRT$. The $GMRT$ is
probably more sensitive than $NRH$ to phase deviations because of its sparser 
$uv$ coverage.
\item
use the shortest possible integration time for $GMRT$ observations ($\approx$ 2
sec., rather than 17 sec. as in the current observations). This is because 
radio bursts can vary on timescales much shorter than 17 sec, and the data 
presented here could have integrated over several potentially interesting 
time-varying phenomena (as was the case for the snapshot at 09:04:04, Aug 27 
2002).
\end{itemize}

This could lead to images with better resolution and dynamic ranges. We could 
then conclusively answer some long-standing questions in solar physics, as 
mentioned in the introduction. 

\begin{acknowledgements}
We thank the staff of the GMRT and the NRH that made these observations 
possible. The GMRT is run by the National Centre for Radio Astrophysics of the 
Tata Institute of Fundamental Research, India. The NRH is operated by the 
Observatoire de Paris, CNRS, R\'{e}gion Centre. We warmly acknowledge 
continued financial support from the French embassy in India without which 
this work could not have been accomplished.
\end{acknowledgements}

\end{document}